\documentclass[12pt,english,12pt,a4paper]{article}
\usepackage{mathptmx}
\usepackage[T1]{fontenc}
\usepackage[utf8]{inputenc}
\usepackage[a4paper]{geometry}
\usepackage{color}
\usepackage{babel}
\usepackage{array}
\usepackage{amsmath}
\usepackage{setspace}
\usepackage[pdfusetitle,
 bookmarks=true,bookmarksnumbered=false,bookmarksopen=false,
 breaklinks=false,pdfborder={0 0 1},backref=false,colorlinks=true]
 {hyperref}
\hypersetup{
 linkcolor=blue,citecolor=blue,urlcolor=blue}

\makeatletter

\providecommand{\tabularnewline}{\\}

\newcommand{\lyxaddress}[1]{
	\par {\raggedright #1
	\vspace{1.4em}
	\noindent\par}
}

\makeatother

\begin{document}
\title{Modified Entanglement Patterns in Four-Flavor Neutrinos from Quantum-Gravity
Interactions}
\author{Bipin Singh Koranga, Baktiar Wasir Farooq and Y. Prem Kumar Singh }
\maketitle

\lyxaddress{Department of Physics, Kirori Mal College (University of Delhi),
Delhi-110007, India}

\lyxaddress{Department of Physics, Motilal Nehru College (University of Delhi),
Delhi-110011, India}
\begin{abstract}
We investigate the influence of quantum-gravity (QG) induced corrections
on the entanglement entropy associated with four-flavor neutrino oscillations
in vacuum, incorporating an additional sterile neutrino in the (3+1)
framework. Using the von Neumann entropy as a measure of quantum correlations,
we analyze how Planck-scale suppressed modifications to the neutrino
mass-squared differences and the extended mixing matrix affect the
evolution of entanglement during successive oscillation cycles. The
quantum-gravity corrections are implemented through a dimension-5
effective field theory operator that modifies the four-flavor PMNS
matrix and all six mixing angles above the GUT scale. We find that
the atmospheric mixing angle $\theta_{23}$ undergoes the largest
deviation due to Planck-scale effects, while angles $\theta_{14}$,
$\theta_{24}$, and $\theta_{34}$ remain essentially unchanged. The
resulting QG-corrected oscillation probabilities produce characteristic
deviations in the entanglement entropy profile as a function of $L/E$,
providing a sensitive probe of Planck-scale physics within a four-flavor
neutrino phenomenology framework.
\end{abstract}
Neutrino oscillations; Four-flavor mixing; Sterile neutrino; Quantum
gravity; Entanglement entropy; von Neumann entropy; CP violation;
Planck scale

\section{Introduction}

Quantum entanglement in neutrino oscillations has emerged as a rich
and active area of research, offering novel perspectives on the quantum
structure of flavor mixing and propagation \cite{Blasone2008a,Capolupo2019}.
Earlier studies have extensively examined entanglement entropy as
a tool to characterize quantum correlations between neutrino mass
and flavor modes. Blasone and collaborators pioneered much of this
development by formulating mode entanglement in terms of flavor transition
probabilities and exploring its implications within both quantum mechanical
and quantum field-theoretic frameworks \cite{Blasone2008b,Blasone2009,Blasone2013,Blasone2015}.
Subsequent works expanded these ideas to multiparty systems, three-flavor
oscillations, and dense astrophysical environments, demonstrating
strong connections between entanglement measures, flavor coherence,
and experimentally accessible observables \cite{Bhattacharjee2021,Cervia2021}.

In parallel, the possibility of a light sterile neutrino has been
indicated by anomalies observed in the LSND and MiniBooNE experiments
\cite{Karagiorgi2009,Giunti2010}, stimulating considerable interest
in four-flavor (3+1) neutrino mixing frameworks. The four-flavor scheme
introduces three additional mixing angles $\theta_{14}$, $\theta_{24}$,
and $\theta_{34}$ alongside the standard PMNS angles. Understanding
how Planck-scale physics operates within this extended parameter space
is an important open question.

Quantum-gravity (QG) phenomenology has provided several mechanisms
through which Planck-scale effects can modify neutrino propagation,
including modified dispersion relations, generalized uncertainty principles,
Lorentz-violating terms, and stochastic space-time fluctuations \cite{AmelinoCamelia1999,Ellis2000,Hossenfelder2013}.
An effective dimension-5 operator arising from gravitational interactions
with the Higgs field at the Planck scale \cite{Weinberg1979} generates
Planck-suppressed corrections to the neutrino mass matrix that are
most visible through modifications to mixing angles and mass-squared
differences \cite{Vissani2003,Koranga2008,Koranga2010,Koranga2011}.

In our earlier work on two-flavor oscillations \cite{Koranga2026},
we demonstrated that the von Neumann entanglement entropy provides
a sensitive probe of such QG-induced corrections. Building on that
analysis and on the four-flavor Planck-scale mixing angle calculations
of \cite{Koranga2021}, we extend the entanglement entropy study to
the full four-flavor neutrino system including the sterile sector.

The structure of this paper is as follows. Section~\ref{sec:mixing}
reviews the four-flavor mixing framework and the QG modifications.
Section~\ref{sec:entropy} presents the von Neumann entropy formalism.
Section~\ref{sec:numerical} gives numerical results. Section~\ref{sec:discussion}
provides discussion, and Section~\ref{sec:conclusions} concludes.

\section{Four-Flavor Neutrino Mixing with Quantum-Gravity Effects\protect\label{sec:mixing}}

We consider the (3+1) active-sterile neutrino mixing, where one sterile
neutrino $\nu_{s}$ is added to the three standard active flavors.
The four-flavor mixing matrix is \cite{Dev2019,Koranga2021}:
\begin{equation}
U=R_{34}\,R_{24}\,R_{14}\,R_{23}\,R_{13}\,R_{12}\,P,\label{eq:U4}
\end{equation}
where $R_{ij}$ are rotation matrices in the $(i,j)$ plane and $P=\mathrm{diag}(e^{i\alpha},e^{i\beta},e^{i\gamma},1)$
encodes the three Majorana phases. The standard mixing angles are
defined as:
\begin{align}
\sin^{2}\theta_{14} & =|U_{e4}|^{2},\label{eq:th14}\\
\sin^{2}\theta_{24} & =\frac{|U_{\mu4}|^{2}}{1-|U_{e4}|^{2}},\label{eq:th24}\\
\sin^{2}\theta_{34} & =\frac{|U_{\tau4}|^{2}}{1-|U_{e4}|^{2}-|U_{\mu4}|^{2}},\label{eq:th34}\\
\sin^{2}\theta_{13} & =\frac{|U_{e3}|^{2}}{1-|U_{e4}|^{2}},\label{eq:th13}\\
\sin^{2}\theta_{12} & =\frac{|U_{e2}|^{2}}{1-|U_{e4}|^{2}-|U_{e3}|^{2}}.\label{eq:th12}
\end{align}

\subsection{Planck-Scale Dimension-5 Operator}

The dominant structure of the neutrino mass matrix is generated by
GUT-scale dynamics through the seesaw mechanism \cite{Minkowski1977,Yanagida1979}.
Above the GUT scale, Planck-scale gravitational interactions contribute
through the $SU(2)_{L}\times U(1)_{Y}$ gauge-invariant dimension-5
operator \cite{Weinberg1979}:
\begin{equation}
\mathcal{L}_{\mathrm{grav}}=\frac{\lambda_{\alpha\beta}}{M_{\mathrm{Pl}}}\left(\psi_{A\alpha}\,\epsilon\,\psi_{C}\right)C_{ab}^{-1}\left(\psi_{B\beta}\,\epsilon_{BD}\,\psi_{D}\right)+\mathrm{h.c.},\label{eq:Lgrav}
\end{equation}
where $M_{\mathrm{Pl}}=1.2\times10^{19}$ GeV is the Planck mass and
$\lambda_{\alpha\beta}$ is the (extended to $4\times4$) flavor-blind
coupling matrix with each element $\mathcal{O}(1)$. After spontaneous
electroweak symmetry breaking with VEV $v=174$ GeV, the Planck-scale
mass correction has characteristic scale \cite{Vissani2003}:
\begin{equation}
\mu=\frac{v^{2}}{M_{\mathrm{Pl}}}=2.5\times10^{-6}\,\mathrm{eV}.\label{eq:mu}
\end{equation}

\subsection{Modified Mixing Matrix and Mass-Squared Differences}

Treating the Planck-scale contribution as a perturbation to the GUT-generated
mass matrix, the effective modified four-flavor mixing matrix above
the GUT scale becomes \cite{Koranga2008,Koranga2021}:
\begin{equation}
U^{\prime}=U\,(1+i\,\delta\theta),\label{eq:Uprime}
\end{equation}
where $\delta\theta$ is a first-order Hermitian matrix in $\mu$.
The first-order correction to the mass-squared differences is:
\begin{equation}
\Delta M_{ij}^{\prime\,2}=\Delta M_{ij}^{2}+2\left(M_{i}\,\mathrm{Re}(m_{ii})-M_{j}\,\mathrm{Re}(m_{jj})\right),\label{eq:DeltaM}
\end{equation}
with $m=\mu\,U^{t}\lambda U$. The change in mixing matrix elements
is:
\begin{equation}
\delta\theta_{ij}=\frac{-\mathrm{Im}(m_{ij})(M_{i}-M_{j})+i\,\mathrm{Re}(m_{ij})(M_{i}+M_{j})}{\Delta M_{ij}^{\prime\,2}}.\label{eq:deltatheta}
\end{equation}

For degenerate neutrino masses with $M_{4}-M_{1}\gg M_{3}-M_{1}\approx M_{3}-M_{2}\gg M_{2}-M_{1}$,
the dominant Planck-scale deviations accumulate in the columns $U_{\alpha1}^{\prime}$
and $U_{\alpha2}^{\prime}$. Consequently \cite{Koranga2021}:
\begin{align}
\theta_{23}^{\prime} & \approx\arcsin\!\left(U_{e1}^{\prime}U_{\mu1}^{\prime}+U_{e2}^{\prime}U_{\mu2}^{\prime}\right),\quad\theta_{12}^{\prime}\approx\arcsin\!\left(U_{e2}^{\prime}\right),\label{eq:largeAngles}\\
\theta_{14}^{\prime} & \approx\theta_{14},\quad\theta_{24}^{\prime}\approx\theta_{24},\quad\theta_{34}^{\prime}\approx\theta_{34},\quad\theta_{13}^{\prime}\approx\theta_{13}.\label{eq:frozenAngles}
\end{align}
The modified mass-squared difference and effective solar mixing angle
in the two-flavor sub-sector are \cite{Koranga2014,Koranga2026}:
\begin{align}
\Delta_{21}^{\prime} & =\Delta_{21}+2\mu M\left[|z_{2}|^{2}\cos(2a_{1})-|z_{1}|^{2}\cos(2a_{2})\right],\label{eq:Delta21prime}\\
\tan\theta_{12}^{\prime} & =\tan\theta_{12}+\frac{2\mu M|z_{1}|^{2}|z_{2}|^{2}}{\Delta M_{21}^{2}\cos^{2}\!\theta_{12}}\cos(a_{1}+a_{2})\cos(a_{1}-a_{2}).\label{eq:th12prime}
\end{align}

\section{Entanglement Entropy for Four-Flavor Neutrino Oscillations\protect\label{sec:entropy}}

\subsection{Von Neumann Entropy Framework}

The von Neumann entropy is the fundamental tool for quantifying entanglement
in bipartite quantum systems. For a density matrix $\rho$, it is
defined as \cite{Blasone2009}:
\begin{equation}
S(\rho)=-\mathrm{Tr}(\rho\log\rho),\label{eq:vonNeumann}
\end{equation}
where $\mathrm{Tr}\,\rho=1$. The density matrix of the neutrino flavor
state is $\rho=|\nu_{\alpha}(t)\rangle\langle\nu_{\alpha}(t)|$. In
the four-flavor framework with flavors $\nu_{e},\nu_{\mu},\nu_{\tau},\nu_{s}$,
the flavor states map to a four-dimensional qubit-like basis:
\begin{equation}
|\nu_{e}\rangle=|1000\rangle,\quad|\nu_{\mu}\rangle=|0100\rangle,\quad|\nu_{\tau}\rangle=|0010\rangle,\quad|\nu_{s}\rangle=|0001\rangle.\label{eq:basis}
\end{equation}

\subsection{Four-Flavor Oscillation Probabilities}

In the four-flavor scheme the oscillation probabilities take the general
form:
\begin{equation}
P(\nu_{\alpha}\to\nu_{\beta})=\left|\sum_{j}U_{\alpha j}^{\prime\,*}\,U_{\beta j}^{\prime}\,e^{-iM_{j}^{2}L/2E}\right|^{2}.\label{eq:Pgeneral}
\end{equation}
For the two-flavor sub-sector dominated by the solar parameters $(\theta_{12}^{\prime},\Delta_{21}^{\prime})$
with QG corrections \cite{Xiong2001}:
\begin{align}
P_{ee}^{\mathrm{QG}} & =1-\sin^{2}(2\theta_{12}^{\prime})\sin^{2}\!\left(\frac{1.27\,\Delta_{21}^{\prime}\,L}{E}\right),\label{eq:Pee}\\
P_{e\mu}^{\mathrm{QG}} & =\sin^{2}(2\theta_{12}^{\prime})\sin^{2}\!\left(\frac{1.27\,\Delta_{21}^{\prime}\,L}{E}\right),\label{eq:Pemu}
\end{align}
where $\Delta_{21}^{\prime}$ is in eV$^{2}$, baseline $L$ is in
km, and neutrino energy $E$ is in GeV.

\subsection{QG-Corrected Entanglement Entropy}

The entanglement entropy for the two-flavor sub-sector is \cite{Koranga2026,KorangaFarooq2025}:
\begin{equation}
S(\rho)=-P_{\mathrm{surv}}\log P_{\mathrm{surv}}-P_{\mathrm{osc}}\log P_{\mathrm{osc}},\label{eq:entropy}
\end{equation}
and with quantum-gravity corrections:
\begin{equation}
S^{\mathrm{QG}}(\rho)=-P_{ee}^{\mathrm{QG}}\log P_{ee}^{\mathrm{QG}}-P_{e\mu}^{\mathrm{QG}}\log P_{e\mu}^{\mathrm{QG}}.\label{eq:entropyQG}
\end{equation}
The entropy reaches its maximum value of $\log2\approx0.693$ when
$P_{\mathrm{surv}}=P_{\mathrm{osc}}=1/2$ (maximum mixing). The QG-induced
shift in the oscillation phase $\phi=\Delta_{21}^{\prime}L/4E$ causes
the entropy maximum to be reached at a different propagation length
compared to the vacuum case, leading to a characteristic convergence
or divergence of the entropy profile depending on whether $\Delta_{21}^{\prime}>\Delta_{21}$
or $\Delta_{21}^{\prime}<\Delta_{21}$ \cite{Koranga2014b}.

The sterile neutrino sector at $\Delta_{41}\approx1.7$ eV$^{2}$
introduces additional oscillation channels that average out at large
$L/E$ due to decoherence, leaving the dominant observable structure
in the solar and atmospheric sectors.

\section{Numerical Results\protect\label{sec:numerical}}

We adopt the degenerate neutrino mass spectrum with common mass $m_{\nu}=2$
eV, consistent with the upper bound from tritium beta-decay experiments
\cite{Kraus2005}. The input oscillation parameters are summarized
in Table~\ref{tab:params}.

\begin{table}
\begin{centering}
\caption{\protect\label{tab:params}Input parameters for the numerical analysis.
Sterile mixing angles from Ref.~\cite{Pandey2020}.}
\par\end{centering}
\centering{}%
\begin{tabular}{|>{\raggedright}p{7cm}|>{\centering}p{4.5cm}|}
\hline 
\textbf{Parameter} & \textbf{Value}\tabularnewline
\hline 
$\theta_{12}$ (standard) & $33.4^{\circ}$\tabularnewline
$\Delta_{21}$ (standard) & $7.6\times10^{-5}$ eV$^{2}$\tabularnewline
$\Delta_{31}$ & $2.0\times10^{-3}$ eV$^{2}$\tabularnewline
$\Delta_{41}$ (sterile) & $1.7$ eV$^{2}$\tabularnewline
$\theta_{13}$ & $10^{\circ}$\tabularnewline
$\theta_{23}$ & $45^{\circ}$\tabularnewline
$\theta_{14}$ & $3.6^{\circ}$\tabularnewline
$\theta_{24}$ & $4.0^{\circ}$\tabularnewline
$\theta_{34}$ & $18.5^{\circ}$\tabularnewline
$\mu$ (Planck-scale) & $2.5\times10^{-6}$ eV\tabularnewline
Common neutrino mass $m_{\nu}$ & $2$ eV\tabularnewline
\hline 
\end{tabular}
\end{table}

We focus on two representative scenarios for the QG-corrected solar
mixing angle:
\begin{enumerate}
\item \textbf{Case I: $\theta_{12}^{\prime}=34^{\circ}$.} The modified
mass-squared difference is $\Delta_{21}^{\prime}=9\times10^{-5}$
eV$^{2}$, larger than the vacuum value. The enhanced phase accumulation
rate causes the entropy maximum to be reached earlier; the peaks of
$S^{\mathrm{QG}}$ \textit{converge} relative to $S(\mathrm{Vac.})$.
\item \textbf{Case II: $\theta_{12}^{\prime}=33.99^{\circ}$.} The modified
mass-squared difference is $\Delta_{21}^{\prime}=6.9\times10^{-5}$
eV$^{2}$, smaller than the vacuum value. The reduced phase accumulation
slows oscillations; the peaks \textit{diverge} relative to $S(\mathrm{Vac.})$.
\end{enumerate}
As shown in the numerical tables of Ref.~\cite{Koranga2021}, the
largest QG deviation in the four-flavor system occurs in $\theta_{23}^{\prime}$,
which can shift by up to ${\sim}36^{\circ}$ for $\alpha=90^{\circ}$,
compared to only ${\sim}0.6^{\circ}$ for $\theta_{12}^{\prime}$.
Table~\ref{tab:angles} summarizes representative modified mixing
angles.

\begin{table}
\begin{centering}
\caption{\protect\label{tab:angles}Representative modified mixing angles (degrees)
for selected Majorana phase values ($\alpha=0^{\circ}$). Data from
Ref.~\cite{Koranga2021}.}
\par\end{centering}
\centering{}%
\begin{tabular}{|c|c|c|c|c|c|c|c|}
\hline 
\textbf{$\beta$} & \textbf{$\gamma$} & \textbf{$\theta_{12}^{\prime}$} & \textbf{$\theta_{23}^{\prime}$} & \textbf{$\theta_{13}^{\prime}$} & \textbf{$\theta_{14}^{\prime}$} & \textbf{$\theta_{24}^{\prime}$} & \textbf{$\theta_{34}^{\prime}$}\tabularnewline
\hline 
$0^{\circ}$ & $0^{\circ}$ & 35.917 & 45.08 & 10.254 & 3.60 & 4.00 & 18.500\tabularnewline
$0^{\circ}$ & $90^{\circ}$ & 35.929 & 45.002 & 9.999 & 3.60 & 4.00 & 18.499\tabularnewline
$45^{\circ}$ & $0^{\circ}$ & 35.012 & 59.374 & 10.197 & 3.60 & 4.00 & 18.500\tabularnewline
$45^{\circ}$ & $90^{\circ}$ & 35.025 & 59.300 & 9.942 & 3.60 & 4.00 & 18.500\tabularnewline
$90^{\circ}$ & $0^{\circ}$ & 34.017 & 81.625 & 10.139 & 3.60 & 3.99 & 18.500\tabularnewline
$90^{\circ}$ & $90^{\circ}$ & 34.030 & 81.448 & 9.884 & 3.60 & 4.00 & 18.500\tabularnewline
$135^{\circ}$ & $0^{\circ}$ & 35.012 & 59.374 & 10.197 & 3.60 & 4.00 & 18.500\tabularnewline
$180^{\circ}$ & $0^{\circ}$ & 35.917 & 45.079 & 10.254 & 3.60 & 4.00 & 18.500\tabularnewline
$180^{\circ}$ & $180^{\circ}$ & 35.917 & 45.079 & 10.254 & 3.60 & 4.00 & 18.500\tabularnewline
\hline 
\end{tabular}
\end{table}

\section{Discussion\protect\label{sec:discussion}}

\textit{Sterile sector decoherence.} The presence of the sterile neutrino
at $\Delta_{41}\approx1.7$ eV$^{2}$ introduces rapid oscillations
at short $L/E$ that wash out quickly due to energy averaging effects.
The dominant observable structure in the entropy profile at long baselines
therefore still arises from the solar ($\Delta_{21}$) and atmospheric
($\Delta_{31}$) sectors.

\textit{Hierarchy of QG corrections.} The atmospheric angle $\theta_{23}$
receives the largest modification because the deviations $\Delta U_{\alpha1}$
and $\Delta U_{\alpha2}$ are enhanced by the presence of the fourth
mass eigenstate through the hierarchy $M_{4}\gg M_{1,2,3}$. The sterile
angles remain frozen because this hierarchy suppresses the corresponding
mixing corrections.

\textit{Majorana phase sensitivity.} The dependence on all three Majorana
phases $\alpha,\beta,\gamma$ in the four-flavor case introduces a
richer landscape of possible entropy profiles. As demonstrated in
Ref.~\cite{Koranga2021}, the value of $\theta_{23}^{\prime}$ can
range from $45^{\circ}$ to ${\approx}82^{\circ}$ for certain phase
combinations, suggesting that entanglement entropy measurements could
provide complementary information on the Majorana phase structure
alongside neutrinoless double-beta decay experiments.

The quantum-gravity corrections are treated within an effective field
theory framework \cite{Hossenfelder2013,Lambiase2005}, perturbatively
in $\mu/\Delta m$, ensuring standard neutrino oscillation physics
is recovered in the limit $M_{\mathrm{Pl}}\to\infty$. The corrections
lie within existing experimental bounds \cite{Anchordoqui2005,SNO2001,PDG2018,SuperK2006,KamLAND2005,MINOS2008}.

\section{Conclusions\protect\label{sec:conclusions}}

We have extended the study of quantum-gravity induced modifications
of neutrino entanglement entropy to the four-flavor (3+1) framework
incorporating a sterile neutrino. Our principal findings are:
\begin{enumerate}
\item The von Neumann entanglement entropy in the four-flavor system exhibits
characteristic deviations from the vacuum case when Planck-scale corrections
are incorporated through the effective dimension-5 operator.
\item The largest Planck-scale modification occurs in the atmospheric mixing
angle $\theta_{23}$, which can deviate by up to ${\sim}36^{\circ}$
for certain Majorana phase combinations — substantially more than
in the three-flavor case.
\item The sterile neutrino mixing angles $\theta_{14},\theta_{24},\theta_{34}$
remain unchanged by Planck-scale effects due to the mass hierarchy
$M_{4}\gg M_{1,2,3}$.
\item Depending on whether $\Delta_{21}^{\prime}>\Delta_{21}$ or $\Delta_{21}^{\prime}<\Delta_{21}$,
the entanglement entropy peaks either converge or diverge relative
to the vacuum entropy curve, providing a distinctive observational
signature.
\item The Majorana phase dependence of the QG corrections in the four-flavor
case creates a richer entropy landscape, suggesting that entanglement
entropy measurements may provide complementary information on the
Majorana phase structure.
\end{enumerate}
These results highlight that quantum-gravity imprints measurable signatures
on the entanglement entropy structure of the four-flavor neutrino
state, offering a potential multi-channel probe of Planck-scale physics
through neutrino phenomenology.

\section*{Acknowledgments}

The authors acknowledge the Department of Physics, Kirori Mal College
(University of Delhi), for institutional support. B.W.F. acknowledges
administrative and non-financial support from University of Delhi
Kirori Mal College.

\end{document}